\documentclass[aps,prd,twocolumn,floatfix,nofootinbib]{revtex4}   

\usepackage{amsmath}    
\usepackage{graphicx}   

\newcommand{\opbraket}[3]{\langle #1|#2|#3\rangle}
\begin{document}

\title{Classical mechanics as the many-particle limit of quantum mechanics}
\author{Gabriele Carcassi}
 \affiliation{Brookhaven National Laboratory, Upton, NY 11973}
 \email{carcassi@bnl.gov}
\date{February 1, 2009}

\begin{abstract}
We derive the classical limit of quantum mechanics by describing the center of mass of a system constituted by a large number of particles. We will show that in that limit the commutator between the position and velocity of the center of mass is infinitesimal, which allows both to be known with great precision. We then show how the infinitesimal commutator allows for the definition of functions of position and velocity, and how the commutator reduces to a Poisson bracket.
\end{abstract}

\maketitle

\section{Introduction}

In contrast to the consistent way in which textbooks present the classical limit of special relativity as the limit for velocities much smaller than the speed of light, the classical limit for quantum mechanics is discussed in many different ways. Some books concentrate on a concrete physical example and show how this behaves classically under some condition \cite{feynman,shankar,liboff} (e.g.~particle moving in a potential with slow and smooth variations, Compton wavelength or spreading of a gaussian wavepacket for masses at our scale) while others state how, mathematically, one can go from one theory to the other \cite{sakurai} (e.g.~limit of small $\hbar$, substitution of commutators with Poisson brackets). The advantage of the first approach is that the physical reasoning in the specific case is clear, but it may not be obvious how to generalize or how the mathematical framework itself changes from quantum to classical. The situation with the second approach is reversed: we may see how it works in the general case or how the mathematical framework changes, but the physical understanding is somewhat lost. It is also interesting to note some textbooks avoid the matter and settle for a comparison between the two theories \cite{sudbery}.

In this paper we offer an alternative way of presenting and deriving the classical limit for quantum mechanics that we hope takes the best of those two approaches. We start by studying the position and velocity of the center of mass of a system in the limit of a large number of particles, show that in this limit their commutator is infinitesimal and they can both be known accurately. This result and the reasoning that follows is similar to what is found in many textbook that follow the first strategy, so one can integrate this work with other material. We then go on showing how, in this limit, the expectation for functions of position and velocity is the function of the expectation of the two, and how the commutator between those functions reduces to Poisson brackets. This gives us the advantages of the texts that follow the second strategy; it shows how the mathematical framework changes, but the novelty is that we are able to derive those fundamental relationships from our case study. Moreover, we are able to do so with mathematical techniques which are comparable to what is already found in quantum textbooks.

\section{Center of mass}

Any system we study in classical mechanics is made of a large number of elementary particles. When we talk about position and velocity of that body, we typically refer to the position and velocity of the center of mass. Let us study, then, the characteristics of position and velocity of the center of mass of a large number of particles in the context of quantum mechanics.

The Hilbert space we need to consider is the product of the Hilbert spaces that describe position and momentum of $N$ particles. To simplify our discussion, we consider only position and momentum along one direction. For each particle we will have its mass $m_i$, its position operator $X_i$ and its momentum operator $P_i$. The commutation relationship between position and momentum operators will be
\begin{align*}
[ X_i, P_j ] &= i\hbar\delta_{ij}
\end{align*}
meaning that they commute if and only if they refer to different particles.

The total mass of the system is given by
\begin{align*}
M &= \displaystyle\sum_{i=1}^N m_i = N \overline{m}
\end{align*}
where $\overline{m}$ is the average particle mass.
The position and velocity of the center of mass are given by
\begin{align*}
X_{CM} = \displaystyle\sum_{i=1}^N \frac{m_i X_{i}}{M}
&&V_{CM} = \displaystyle\sum_{i=1}^N \frac{P_{i}}{M}
\end{align*}
The commutator between them is given by
\begin{align*}
[ X_{CM}, V_{CM}] &= \displaystyle\sum_{i=1}^N \displaystyle\sum_{j=1}^N \frac{m_i}{M^2} [ X_i, P_j ]
= \displaystyle\sum_{i=1}^N \frac{i\hbar m_i}{M^2}
= \frac{i\hbar}{N\overline{m}}
\end{align*}
If we take the limit of large $N$, the commutator tends to 0. We recall that the commutation relationship is what sets the limit on the uncertainty, so what this means is that the more particles we have, the more precisely the position and velocity of the center of mass can be known. Note that the uncertainty on position and momentum for each particle is the same, while \emph{at the same time} the uncertainty for $X_{CM}$ and $V_{CM}$ is much smaller. In other words: it is possible to know the position and velocity of the center of mass much more precisely than the position and momentum of each individual particle.

Given a certain level of accuracy needed to give a satisfactory description of the system, we can use the uncertainty relation
\begin{align*}
\Delta x_{CM}\Delta v_{CM} \geq \frac{\hbar}{2N\overline{m}}
\end{align*}
to determine when we can consider $N$ large enough to ignore quantum effects. An effective way is to compare the quantum uncertainty with other sources of uncertainty and see what dominates. For example, when we think of a baseball we might imagine a perfect homogeneous sphere with the center of mass positioned perfectly at the center. In practice, though, a baseball is not an ideal sphere and the mass will be only known within a certain margin of error. This gives a systematic uncertainty on the position and velocity of the center of mass, which can be quite large compared to the quantum uncertainty. If we are describing an electron going through a double slit, its mass too is known only within a certain margin of error, but the related systematic uncertainty will be smaller than the quantum uncertainty. The quantum uncertainty will still exist for bodies consisting of a large number of particles but in practice will be disregarded because other sources of uncertainty are going to be larger; in the same way relativistic effects exist also for slow moving bodies but in practice can be ignored as they would be hidden by the uncertainty of our measurements.

A few other items are of note.

First is that in the same Hilbert space we have observables that exhibit quantum behavior, such as the position and momentum of each individual particle, and observables that exhibit classical behavior, such as the position and velocity of the center of mass. This is important because it shows how quantum and classical mechanics co-exist, which is not the case if we take the limit of small $\hbar$ or if we substitute commutators with Poisson brackets. In those cases it is more like giving a recipe to go from one theory to another.

Second, we note that if we let $\hbar$ go to zero we have the same effect on the commutator, so our work is not altogether different. While mathematically it may be the same, though, we believe that this approach is physically less precise. First $\hbar$ is a constant, and as such does not change and should be the same in all physical accounts. Second it is not clear what $\hbar$ physically represents, so it is even less clear what the limit means. Since in this derivation it is the number of particles that accounts for the infinitesimal commutator, the physical meaning of the limit is clear and $\hbar$ can be left at its constant value.

Third, if we consider the total momentum, instead of the velocity of the center of mass, the commutation relationship becomes:
\begin{align*}
[ X_{CM}, P_{TOT}] = i\hbar
\end{align*}
which seems to say that the uncertainty is the same for the individual particle and for the center of mass. The expectation of $P_{TOT}$, though, increases proportionally to $N$, so while the uncertainty is the same in absolute terms, is still going to be decreasing in relative terms. It is precisely to avoid this confusion that we use the position and velocity: both their expectations remain finite as $N$ increases.

Fourth, we write explicitly the total mass in terms of the number of particles and the average mass. We do this to make it clear that the limit is reached by increasing the number of particles, which is what physically happens, rather than by increasing the mass of a fixed number of particles. This also allows one to consider the classical limit for a system composed of massless particles in the following way. Instead of the position and the velocity of the center of mass, we can consider the average position and average momentum, which are quantities defined for both massive and massless particles; by proceeding with the same exact reasoning one finds that the average position and the average momentum can be known to a better precision than the position and momentum of each individual particle. This substitution can be applied to the subsequent arguments and derivations, excluding the section about the equation of motion.

Fifth, our discussion tells us that in the limit of large $N$ we will have states in which both the position and the velocity of the center of mass are well determined. It does not say that these quantities are well determined for all states: a linear combination of states with different values will give us a distribution. In general the system we measure will be prepared according to a distribution, but this is true for classical mechanics as well (we rarely have the ability to prepare a system in an infinitely precise state; initial conditions are known up to some level of accuracy). The important point is that we can select, through measurement, states in which both are well determined because states in which position and velocity are defined but differ in value are orthogonal to each other. We also need to realize that there are an infinite number of states for which the position and the velocity of the center of mass are going to be the same: one for each different configuration of the individual particles. This is indeed what one would expect.

Sixth, the idea that the position of the center of mass can be known to a better precision than each individual particle might seem counterintuitive at first. We note that, in practice, the most common way to improve the precision of a measurement is to repeat it and take the average of the result. If we think that measuring the center of mass is effectively measuring an average of the position of a number of particles this apparent puzzle is resolved.

From now on, we will set
\begin{align*}
\varepsilon = \frac{1}{N\overline{m}} &&
[ X_{CM}, V_{CM}] = i\hbar\varepsilon
\end{align*}
merely for notational convenience.

\section{Functions}

We turn our attention to observables that are a function of $X_{CM}$ and $V_{CM}$. In quantum mechanics we often have problems defining functions in $X$ and $P$: since they do not commute even the definition of a polynomial is ambiguous. Given the infinitesimal commutator between $X_{CM}$ and $V_{CM}$, we can always rearrange the order of a polynomial defined in those operators by adding terms that are of the order of $\varepsilon$. Therefore we can then always write:
\begin{align*}
F(X_{CM}, V_{CM}) &= \displaystyle\sum_{i=0}^\infty \displaystyle\sum_{j=0}^\infty c_{i,j} X_{CM}^i V_{CM}^j + \mathcal{O}(\varepsilon) \\
&= f(X_{CM}, V_{CM}) + \mathcal{O}(\varepsilon)
\end{align*}
where $F$ is defined with a precise order of multiplication while $f$ is not.

We also note that if two operators commute, the expectation value of the product is the product of the expectation value. Given that $X_{CM}$ and $V_{CM}$ commute when $\varepsilon$ tends to 0, we have:
\begin{align*}
\opbraket{\Psi}{X_{CM} V_{CM}}{\Psi} &= \opbraket{\Psi}{X_{CM}}{\Psi}\opbraket{\Psi}{V_{CM}}{\Psi} + \mathcal{O}(\varepsilon) \\
&= x_{CM} \cdot v_{CM} + \mathcal{O}(\varepsilon) \\
\opbraket{\Psi}{F(X_{CM}, V_{CM})}{\Psi} &= f(x_{CM}, v_{CM}) + \mathcal{O}(\varepsilon)
\end{align*}
where
\begin{align*}
x_{CM}&=\opbraket{\Psi}{X_{CM}}{\Psi} \\
v_{CM}&=\opbraket{\Psi}{V_{CM}}{\Psi}
\end{align*}

This means that, in the limit of large number of particles, we can define functions of $X_{CM}$ and $V_{CM}$, which is what we do in classical mechanics.

\section{Commutators}

We now want to calculate the commutator between two functions of $X_{CM}$ and $V_{CM}$. We note that this commutator will also be infinitesimal, as it is a difference between different reordering of polynomials in $X_{CM}$ and $V_{CM}$. So what we are looking for is the first order in $\varepsilon$. As a first step, we study the commutation between powers of $X_{CM}$ and powers of  $V_{CM}$. We have:
\begin{align*}
&[ X_{CM}^n, V_{CM} ] = \displaystyle\sum_{j=0}^{n-1} X_{CM}^{(n-1-j)} [ X_{CM}, V_{CM}] X_{CM}^j \\
&[ X_{CM}^n, V_{CM}^m ] = \displaystyle\sum_{i=0}^{m-1} V_{CM}^{(m-1-i)} [ X_{CM}^n, V_{CM}] V_{CM}^i \\
&= \displaystyle\sum_{i=0}^{m-1} \displaystyle\sum_{j=0}^{n-1} V_{CM}^{(m-1-i)} X_{CM}^{(n-1-j)} [ X_{CM}, V_{CM}] X_{CM}^j V_{CM}^i \\
&= \frac{1}{i\hbar\varepsilon} (\displaystyle\sum_{i=0}^{m-1} V_{CM}^{(m-1-i)} [ X_{CM}, V_{CM}] V_{CM}^i )  \\
&\cdot (\displaystyle\sum_{j=0}^{n-1} X_{CM}^{(n-1-j)} [ X_{CM}, V_{CM}] X_{CM}^j ) + \mathcal{O}(\varepsilon^2)
\end{align*}
where we divided and multiplied by $[X_{CM}, V_{CM}]$ and rearranged the products. Note that the first term is still in the first order in $\epsilon$. We can rewrite as:
\begin{equation}
\label{powercommute}
[ X_{CM}^n, V_{CM}^m ]
= \frac{[ X_{CM}^n, V_{CM} ] [ X_{CM}, V_{CM}^m ]}{i\hbar\varepsilon} + \mathcal{O}(\varepsilon^2)
\end{equation}

We use this relationship to calculate the commutator between polynomials of $X_{CM}$ and $V_{CM}$. We have:
\begin{align*}
&[ X_{CM}^aV_{CM}^b, X_{CM}^cV_{CM}^d ] \\
&= X_{CM}^a [ V_{CM}^b, X_{CM}^c ] V_{CM}^d + X_{CM}^a X_{CM}^c [ V_{CM}^b, V_{CM}^d ] \\
&+ [ X_{CM}^a, X_{CM}^c ] V_{CM}^d V_{CM}^b + X_{CM}^c [ X_{CM}^a, V_{CM}^d ] V_{CM}^b \\
&= X_{CM}^c [ X_{CM}^a, V_{CM} ] [ X_{CM}, V_{CM}^d ] V_{CM}^b \\
&- X_{CM}^a [ X_{CM}^c, V_{CM} ] [ X_{CM}, V_{CM}^b ] V_{CM}^d + \mathcal{O}(\varepsilon^2)
\end{align*}
Changing once again the order of the products, we can write:
\begin{align}
\label{polycommute}
&[ X_{CM}^aV_{CM}^b, X_{CM}^cV_{CM}^d ] \notag \\
&= \frac{[ X_{CM}^a V_{CM}^b, V_{CM} ] [ X_{CM}, X_{CM}^c V_{CM}^d ]}{i\hbar\varepsilon}  \notag \\
&- \frac{[ X_{CM}^c V_{CM}^d, V_{CM} ] [ X_{CM}, X_{CM}^a V_{CM}^b ]}{i\hbar\varepsilon} + \mathcal{O}(\varepsilon^2)
\end{align}

Assuming $f$ and $g$ are Taylor expansible functions and using the linearity of the commutator, we can write:
\begin{align*}
&[f(X_{CM}, V_{CM}), g(X_{CM}, V_{CM})] \\
&= \frac{[ f(X_{CM}, V_{CM}), V_{CM} ] [ X_{CM}, g(X_{CM}, V_{CM}) ]}{i\hbar\varepsilon}  \\
&- \frac{[ g(X_{CM}, V_{CM}), V_{CM} ] [ X_{CM}, f(X_{CM}, V_{CM}) ]}{i\hbar\varepsilon} + \mathcal{O}(\varepsilon^2)
\end{align*}
Using the following relationships:
\begin{align*}
[ f(X_{CM}), V_{CM} ] &= i\hbar\varepsilon \frac{\partial f}{\partial X_{CM}} \\
[ X_{CM}, f(V_{CM}) ] &= i\hbar\varepsilon \frac{\partial f}{\partial V_{CM}} \\
[ f(X_{CM}, V_{CM}), V_{CM} ] &= i\hbar\varepsilon \frac{\partial f}{\partial X_{CM}} + \mathcal{O}(\varepsilon)\\
[ X_{CM}, f(X_{CM}, V_{CM}) ] &= i\hbar\varepsilon \frac{\partial f}{\partial V_{CM}} + \mathcal{O}(\varepsilon)
\end{align*}
we have:
\begin{align}
&\frac{[f(X_{CM}, V_{CM}), g(X_{CM}, V_{CM})]}{i\hbar} \notag \\
&= \varepsilon \left\{\frac{\partial f}{\partial X_{CM}} \frac{\partial g}{\partial V_{CM}} - \frac{\partial g}{\partial X_{CM}} \frac{\partial f}{\partial V_{CM}}\right\} + \mathcal{O}(\varepsilon^2)
\end{align}

\section{Equations of motion}

We will now assume that the Hamiltonian for the center of mass can be written as a function of its position and velocity. Under that assumption, we can study the evolution of an observable using the Heisenberg equation of motion:
\begin{align*}
\frac{d}{dt}f(X_{CM}, V_{CM}) &= \frac{[ f, H ]}{i\hbar}  \\
\frac{d}{dt}f(X_{CM}, V_{CM}) &\approx \varepsilon \left\{\frac{\partial f}{\partial X_{CM}} \frac{\partial H}{\partial V_{CM}} - \frac{\partial H}{\partial X_{CM}} \frac{\partial f}{\partial V_{CM}}\right\}
\end{align*}
Since we usually write functions of total momentum instead of the velocity of the center of mass, we change variable and redefine the functions accordingly:
\begin{align*}
P_{TOT} = N \overline{m} V_{CM} = \frac{1}{\varepsilon} V_{CM}
\end{align*}
\begin{align*}
&\frac{d}{dt} f(X_{CM}, P_{TOT}) \\
&\approx \varepsilon \left\{\frac{\partial f}{\partial X_{CM}} \frac{\partial H}{\partial P_{TOT}} \frac{\partial P_{TOT}}{\partial V_{CM}} - \frac{\partial H}{\partial X_{CM}} \frac{\partial f}{\partial P_{TOT}} \frac{\partial P_{TOT}}{\partial V_{CM}}\right\}
\end{align*}
\begin{align*}
\frac{d}{dt} f(X_{CM}, P_{TOT})
&\approx \left\{ f, H \right\}_{Poisson}
\end{align*}

If we take the expectation on both sides
\begin{align}
\frac{d}{dt} f(x_{CM}, p_{TOT})
&\approx \left\{ f, H \right\}_{Poisson}
\end{align}
we have the same relationship but applied to numbers instead of operators. This is the well known result that in the classical limit we substitute commutators with Poisson brackets \cite{sakurai}. The difference is that we have \emph{derived} that relationship, instead of imposing it.

If we look at the evolution of the operators $X_{CM}$ and $V_{CM}$ when we start from a state in which both position and velocity are well defined, the variation of both is a number, not an operator. This means that we will transition to another state in which both are well determined. If we begin with a distribution of $X_{CM}$ and $V_{CM}$ we can then study the evolution of each individual point in phase space independently: the evolution of the distribution will follow the rules of classical probability.

A few points of discussion.

First, note we have not made any special assumption on what physical processes these particles are participating in. We did not say what kind of interaction goes on between the particles, what external forces are acting on them (except for the requirement on the Hamiltonian), or whether or not they are subject to measurements during the evolution. All that we derived comes from the relationship between the obvervables describing the position and velocity of the center of mass.

Second, the fact that in the limit of large $N$, $X_{CM}$ and $V_{CM}$ remain well defined if they start well defined is already found in textbooks when discussing the evolution of a gaussian packet: it is usually noted how the spread of the packet is negligible for bodies with masses comparable to everyday objects. We look at this from a different angle to make it consistent with the rest of the narrative, but the point is the same.

To sum up, this approach shows that classical evolution applies in the limit of observables whose commutator is infinitesimal: it is the infinitesimal commutator that allows us to define functions of those operators, with the expectation of those functions reduced to the expectation of $X_{CM}$ and $P_{TOT}$, and to substitute commutators with Poisson brackets. We believe this more elegant as it reinforces the idea that some observables are evolving according classical mechanics while others \emph{at the same time and in the same Hilbert space} evolve according to quantum mechanics.

\section{Conclusion}
We have seen that we can derive classical mechanics by studying the evolution of the position and the velocity of the center of mass of a system composed of a large number of particles. In that limit the commutator between position and velocity is infinitesimal, making it possible to define functions of these observables and to replace commutators with Poisson brackets. We have also touched on how this presentation relates to discussions found in a few textbooks.

We believe this derivation to be more straightforward than other discussions on the classical limit because of two main reasons. First because we start off with a clear physical description of what we intend to do, which helps our physical understanding, but we still reach the goal of showing how the mathematical framework reduces to classical mechanics. Second, because it shows how classical mechanics is really a subset of quantum mechanics, more specifically the part that describes the evolution of the expectation value for functions of infinitesimally commuting observables.

\end{document}